\documentclass[12pt]{article}
\begin{document}

\title{\bf Teleparallel Version of the Levi-Civita Vacuum Solutions
and their Energy Contents}

\author{M. Sharif \thanks{msharif@math.pu.edu.pk} and M. Jamil Amir
\thanks{mjamil.dgk@gmail.com}\\
Department of Mathematics, University of the Punjab,\\
Quaid-e-Azam Campus, Lahore-54590, Pakistan.}

\date{}

\maketitle

\begin{abstract}
In this paper, we find the teleparallel version of the Levi-Civita
metric and obtain tetrad and the torsion fields. The tensor, vector
and the axial-vector parts of the torsion tensor are evaluated. It
is found that the vector part lies along the radial direction only
while the axial-vector vanishes everywhere because the metric is
diagonal. Further, we use the teleparallel version of
M$\ddot{o}$ller, Einstein, Landau-Lifshitz and Bergmann-Thomson
prescriptions to find the energy-momentum distribution of this
metric and compare the results with those already found in General
Relativity. It is worth mentioning here that momentum is constant in
both the theories for all the prescriptions. The energy in
teleparallel theory is equal to the corresponding energy in GR only
in M$\ddot{o}$ller prescription for the remaining prescriptions, the
energy do not agree in both theories. We also conclude that
M$\ddot{o}$ller's energy-momentum distribution is independent of the
coupling constant $\lambda$ in the teleparallel theory.

\end{abstract}


\section{Introduction}

Ever since the Einstein's theory of General Relativity (GR) was
proposed, the subject of the localization of energy has been lacking
of a definite answer. It is well-known that the definition of energy
is an oldest, thorny, most interesting and most controversial
problem in GR [1]. Many researchers have proposed the several
energy-momentum complexes to resolve this problem. As a pioneer,
Einstein [2] proposed an expression for the energy-momentum density
of the gravitational field. In fact, this quantity is a
pseudo-tensor, i.e., it is a coordinate dependent object. Later on,
Landau-Lifshitz [3], Papapetrou [4], Bergmann [5], Tolman [6],
Weinberg [7] and M$\ddot{o}$ller [8] proposed their own
prescriptions to resolve this issue. These prescriptions, except
M$\ddot{o}$ller's [8], are restricted to perform the calculations in
Cartesian coordinates.

Virbhadra and his collaborators re-opened this issue and showed that
several prescriptions could give the same results for a given
spacetime [9-14]. He also found that, for a general non-static
spherically symmetric metric of the Kerr-Schild class, the four
different complexes ELLPW (Einstein, Landau-Lifshitz, Papapetrou and
Weinberg) yield the same result as found by [15,16] in the context
of quasi-local mass. However, some other people [17,18] found that
different energy-momentum complexes might give different results for
a given spacetime.

Alternate representations of a theory are usually important and
reflect a valuable insight. The notion of tetrad field was first
introduced by Einstein [19] to unify gravitation and
electromagnetism, besides that he could not succeed. Later on,
Hayashi nad Nakano [20] formulated the tetrad theory of
gravitation, also called teleparallel gravity (TPG) or \textit{New
General Relativity} which corresponds to a gauge theory of
translation group [21,22]. The basic entities of this theory are
the non-trivial tetrad fields ${h^a}_\mu$ and is  defined on
Weitzenb$\ddot{o}$ck spacetime [23], which is endowed with the
affine connection
${\Gamma^\theta}_{\mu\nu}={{h_a}^\theta}\partial_\nu{h^a}_\mu$,
called Weitzenb$\ddot{o}$ck connection. The curvature tensor,
constructed out of this connection, vanishes identically but
torsion remains non-zero.

In TPG, gravitation is attributed to torsion [22] which plays the
role of force [24] while it geometrizes the underlying spacetime in
the case of GR. The translational gauge potentials appear as a
non-trivial part of the tetrad field and induce a teleparallel (TP)
structure on spacetime which is directly related to the presence of
a gravitational field. In some other theories [21-25], torsion is
only relevant when spins are important [26]. This point of view
indicates that torsion might represent some additional degrees of
freedom as compared to curvature and some new physics may be
associated with it. TP is naturally formulated by gauging external
(spacetime) translations which are closely related to the group of
general coordinate transformations underlying GR. Thus the
energy-momentum tensor represents the matter source in the field
equations of tetradic theories of gravity like in GR.

Some authors [27,28] hoped that the problem of localization of
energy might be settled in the framework of TPG and the results may
coincide with those already found in the context of GR. Mikhail et
al. [27] gave the TP version of M$\ddot{o}$ller prescription and
Vargas [29] constructed the TP version of Einstein, Landau-Lifshitz
and Bergmann-Thomson prescriptions. Using the TP version of Einstein
and Landau-Lifshitz prescriptions, Vargas showed that total energy
of the closed FRW universe is zero which agrees with the results
obtained by Rosen [30]. After this, many authors [31] explored the
energy-momentum distribution of different spacetimes by using the TP
version of the above mentioned prescriptions. It is found that the
results are same in both the theories for some spacetimes while they
disagree in some cases.

Pereira, et al. [32] obtained the TP versions of the Schwarzschild
and the stationary axisymmetric Kerr solutions of GR. They proved
that the axial-vector torsion plays the role of the
gravitomagnetic component of the gravitational field in the case
of slow rotation and weak field approximations. Recently [33], we
have found the TP versions of the Friedmann models and
Lewis-Papapetrou spacetimes which lead to some interesting
results. We have also explored the energy-momentum distribution of
the Lewis-Papapetrou spacetime by using TP version of
M$\ddot{o}$ller prescription [34]. It has been extended to the
stationary axisymmetric solutions of Einstein-Maxwell field
equations [35] and the class of static axially symmetric solutions
of EFEs [36] together with their energy contents. The irreducible
parts of the torsion tensor and the axial-vectors are also
investigated.

This paper is devoted to find the TP version of the Levi-Civita
vacuum solutions and then calculate the irreducible parts of the
torsion tensor. The energy-momentum distribution of the solutions is
also explored by using the TP version of M$\ddot{o}$ller, Einstein,
Landau-Lifshitz and Bergmann-Thomson prescriptions. The results for
energy contents are compared with those found in the framework of GR
[18]. The description of this paper is as follows. Section $2$
contains the review of the basic concepts of the TP theory and the
TP version of M$\ddot{o}$ller, Einstein, Landau-Lifshitz and
Bergmann-Thomson prescriptions. In section $3$, we shall find the TP
version of the Levi-Civita vacuum solutions and the irreducible
parts of the torsion tensor. Section $4$ is devoted to the
evaluation of the energy-momentum distribution for Levi-Civita
metric using these prescriptions. The last section provides summary
and discussion of the results obtained.

\section{Teleparallel Theory and Energy-Momentum Prescriptions}

The basic entity of the theory of TPG is the non-trivial tetrad
[37] ${h^a}_\mu$ whose inverse is denoted by ${h_a}^\nu$ and
satisfy the following relations:
\begin{equation}
{h^a}_\mu{h_a}^\nu={\delta_\mu}^\nu; \quad\
{h^a}_\mu{h_b}^\mu={\delta^a}_b.
\end{equation}
The theory of TPG is described by the Weitzenb$\ddot{o}$ck
connection given by
\begin{equation}
{\Gamma^\theta}_{\mu\nu}={{h_a}^\theta}\partial_\nu{h^a}_\mu
\end{equation}
which comes from the condition of absolute parallelism [22]. This
implies that the spacetime structure underlying a translational
gauge theory is naturally endowed with a TP structure [21,22]. In
this paper, if not mentioned specifically, the Latin alphabet
$(a,b,c,...=0,1,2,3)$ will be used to denote the tangent space
indices and the Greek alphabet $(\mu,\nu,\rho,...=0,1,2,3)$ to
denote the spacetime indices. The Riemannian metric in TPG arises as
a by product [22] of the tetrad field given by
\begin{equation}
g_{\mu\nu}=\eta_{ab}{h^a}_\mu{h^b}_\nu,
\end{equation}
where $\eta_{ab}$ is the Minkowski spacetime such that
$\eta_{ab}=diag(+1,-1,-1,-1)$. For the Weitzenb$\ddot{o}$ck
spacetime, the torsion is defined as [38]
\begin{equation}
{T^\theta}_{\mu\nu}={\Gamma^\theta}_{\nu\mu}-{\Gamma^\theta}_{\mu\nu}
\end{equation}
which is antisymmetric w.r.t. its last two indices. Due to the
requirement of absolute parallelism, the curvature of the
Weitzenb$\ddot{o}$ck connection vanishes identically [37]. The
Weitzenb$\ddot{o}$ck connection and the Christoffel symbol satisfy
the following relation
\begin{equation}
{{\Gamma^{0}}^\theta}_{\mu\nu}={\Gamma^\theta}_{\mu\nu}
-{K^{\theta}}_{\mu\nu},
\end{equation}
where ${{\Gamma^{0}}^\theta}_{\mu\nu} $ are the Christoffel symbols
and ${K^{\theta}}_{\mu\nu}$ denotes the {\bf contorsion tensor}
given as
\begin{equation}
{K^\theta}_{\mu\nu}=\frac{1}{2}[{{T_\mu}^\theta}_\nu+{{T_\nu}^
\theta}_\mu-{T^\theta}_{\mu\nu}].
\end{equation}

The torsion tensor of the Weitzenb$\ddot{o}$ck connection can be
decomposed into three irreducible parts under the group of global
Lorentz transformations [22]: the tensor part
\begin{equation}
t_{\lambda\mu\nu}={\frac{1}{2}}(T_{\lambda\mu\nu}
+T_{\mu\lambda\nu})+{\frac{1}{6}}(g_{\nu\lambda}V_\mu
+g_{\nu\mu}V_\lambda)-{\frac{1}{3}}g_{\lambda\mu}V_\nu,
\end{equation}
the vector part
\begin{equation}
{V_\mu}={T^\nu}_{\nu\mu}
\end{equation}
and the axial-vector part
\begin{equation}
{A^\mu}=\frac{1}{6}\epsilon^{\mu\nu\rho\sigma} T_{\nu\rho\sigma},
\end{equation}
where
\begin{equation}
\epsilon^{\lambda\mu\nu\rho}= \frac{1}{\surd{-g}}
\delta^{\lambda\mu\nu\rho}.
\end{equation}
Here $\delta=\{\delta^{\lambda\mu\nu\rho}\}$ and
$\delta^*=\{\delta_{\lambda\mu\nu\rho}\}$ are completely skew
symmetric tensor densities of weight -1 and +1 respectively [22].

The TP version of the Einstein, Landau-Lifshitz and
Bergman-Thomson energy-momentum complexes, by setting $c=1=G$, are
respectively given by [29]
\begin{eqnarray}
hE_\nu^\mu&=&\frac{1}{4\pi}\partial_\lambda({U_\nu}^{\mu\lambda}),\\
hL^{\mu\nu}&=&\frac{1}{4\pi}\partial_\lambda(hg^{\mu\beta}{U_\beta}
^{\nu\lambda}),\\
hB^{\mu\nu}&=&\frac{1}{4\pi}\partial_\lambda({g^{\mu\beta}U_\beta}
^{\nu\lambda}),
\end{eqnarray}
where ${U_\nu}^{\mu\lambda}$ is the  Freud's superpotential and is
given dy
\begin{equation}
{U_\nu}^{\mu\lambda}=h{S_\nu}^{\mu\lambda}.
\end{equation}
Here $S^{\nu\mu\lambda}$ is a  tensor quantity, which is skew
symmetric in its last two indices, and is defined as
\begin{equation}
S^{\mu\nu\lambda}=m_1T^{\nu\mu\lambda}+\frac{m_2}{2}(T^{\mu\nu\lambda}-
T^{\lambda\nu\mu})+\frac{m_3}{2}(g^{\nu\lambda}{T^{\beta\mu}}_
\beta-g^{\mu\nu} {T^{\beta\lambda}}_\beta),
\end{equation}
where $m_1$, $m_2$ and $m_3$ are three dimensionless coupling
constants of TPG [22]. TPG equivalent of GR may be obtained by
considering the following particular choice
\begin{equation}
m_1=\frac{1}{4}, \quad\quad m_2=\frac{1}{2}, \quad\quad m_3=-1
\end{equation}
It is mentioned here that $hE_0^0$, $hL^{00}$, $hB^{00}$ are the
energy density components and $hE_i^0$, $hL^{0i}$, $hB^{0i}, (i=1,
2, 3)$ are the momentum density components of the Einstein,
Landau-Lifshitz and Bergman-Thomson prescriptions respectively.

The superpotential of the M$\ddot{o}$ller tetrad theory is given
by Mikhail et al. [27] as
\begin{equation}
{U_\mu}^{\nu\beta}=\frac{\sqrt{-g}}{2\kappa}P_{\chi\rho\sigma}^{\tau\nu\beta}
[{V^\rho}g^{\sigma\chi} g_{\mu\tau}-\lambda g_{\tau\mu}
K^{\chi\rho\sigma}-g_{\tau\mu}(1-2\lambda) K^{\sigma\rho\chi}],
\end{equation}
where
\begin{equation}
P_{\chi\rho\sigma}^{\tau\nu\beta}= {\delta_\chi}^{\tau}
g_{\rho\sigma}^{\nu\beta}+{\delta_\rho}^{\tau}
g_{\sigma\chi}^{\nu\beta}-{\delta_\sigma}^{\tau}
g_{\chi\rho}^{\nu\beta},
\end{equation}
while $ g_{\rho\sigma}^{\nu\beta}$ is the tensor quantity and is
defined by
\begin{equation}
g_{\rho\sigma}^{\nu\beta}={\delta_\rho}^{\nu}{\delta_\sigma}^{\beta}-
{\delta_\sigma}^{\nu}{\delta_\rho}^{\beta},
\end{equation}
$K^{\sigma\rho\chi}$ is the contortion tensor, $g$ is the
determinant of the metric tensor $g_{\mu\nu},~\lambda$ is free
dimensionless coupling constant of TPG, $\kappa$ is the Einstein
constant and $V_\mu$ is the basic vector field. We can write the
M$\ddot{o}$ller energy-momentum density components as
\begin{equation}
\Xi_\mu^\nu= U_\mu^{\nu\rho},_\rho,
\end{equation}
where comma means ordinary differentiation. Here $\Xi_0^0$ and
$\Xi_i^0,~(i=1, 2, 3)$ are the energy and momentum density
components respectively.

\section{Teleparallel Version of the Levi-Civita Solutions}

In 1917, Levi-Civita obtained the most general static
cylindrically symmetric vacuum solutions [39]. Since then, many
explicit cylindrical solutions, for different fluids, have been
found. These solutions are mostly local and no global analysis has
been usually discussed. Recently, Bicak et al. [40] have studied
the global properties of static cylindrically symmetric spacetimes
with perfect fluid matter. The global existence of these solutions
and the finiteness of radius of fluid cylinder is shown. It is
also shown that when the fluid cylinder has a finite extension, it
is possible to glue it smoothly with Levi-Civita solutions and
obtain a global solution. The metric is given by [41]
\begin{equation}
ds^2=\rho^{4s}dt^2-\rho^{4s(2s-1)}(d\rho^2+dz^2)-\alpha^2\rho^{2(1-2s)}d\phi^2,
\end{equation}
where $\alpha$ is a parameter and $s$ is the charge density parameter.
The following interpretations are somewhat expected for:\\
$s=0,~\alpha=1/2$, the above metric reduces to locally flat
spacetime\\
$s=0,~\alpha=1$, it represents the Minkowski spacetime\\
$s=0,~\alpha\neq1$, we have cosmic string.\\
It is mentioned here that $\gamma$-metric is one of the most
interesting metrics of the family of the Weyl solutions. It is
also known as Zipoy-Voorhes metric [42]. The Levi-Civita metric
can be obtained from the $\gamma$-metric in the limiting case when
the length of its Newtonian image source tends to infinity.

The tetrad field satisfying Eq.(3) is given as

\begin{equation}
{h^a}_\mu=\left\lbrack\matrix {\rho^{2s}  &&&   0    &&&   0 &&& 0
\cr 0        &&& \rho^{2s(2s-1)} &&&      &&&   0 \cr 0 &&& 0 &&&
\alpha\rho^{(1-2s)}&&& 0 \cr 0        &&&   0 &&& 0 &&&
\rho^{2s(2s-1)} \cr } \right\rbrack
\end{equation}
with its inverse
\begin{equation}
{h_a}^\mu=\left\lbrack\matrix {\rho^{-2s}  &&&   0    &&&   0 &&& 0
\cr 0        &&& \rho^{2s(1-2s)} &&&      &&&   0 \cr 0 &&& 0 &&&
\frac{1}{\alpha}\rho^{2s-1}&&& 0 \cr 0        &&&   0 &&& 0 &&&
\rho^{2s(1-2s)} \cr} \right\rbrack.
\end{equation}
The non-vanishing components of the Weitzenb$\ddot{o}$ck connection
can be found by using Eqs.(22) and (23) in Eq.(2) as
\begin{eqnarray} {\Gamma^0}_{01}&=&\frac{2s}{\rho}, \quad
{\Gamma^1}_{11}=\frac{2s(2s-1)}{\rho},\nonumber\\
{\Gamma^2}_{21}&=&\frac{1-2s}{\rho},\quad\
{\Gamma^3}_{31}=\frac{2s(2s-1)}{\rho}.
\end{eqnarray}
The corresponding non-vanishing components of the torsion tensor
are
\begin{eqnarray}
{T^0}_{01}&=& -\frac{2s}{\rho}=-{T^0}_{10},\quad\ {T^2}_{21}=
\frac{2s-1}{\rho}=-{T^2}_{12},\nonumber\\
\quad\ {T^3}_{31}&=&\frac{2s(1-2s)}{\rho}=-{T^3}_{13}.
\end{eqnarray}
When we make use of these values in Eqs.(7)-(9), we get the
following non-vanishing components of the tensor part
\begin{eqnarray}
t_{010}&=&\frac{-1}{6}(4s^2-8s+1)\rho^{4s-1}=t_{100},\quad\
t_{001}=-2t_{001}, \nonumber\\
t_{212}&=&\frac{\alpha^2}{3}(2s^2+2s-1)\rho^{1-4s}=t_{122},\quad\
t_{221}=-2t_{212}, \nonumber\\
t_{313}&=&\frac{-1}{6}(8s^2-4s-1)\rho^{8s^2-4s-1}=t_{133},\quad
t_{331}=-2t_{313}
\end{eqnarray}
and the vector part
\begin{equation}
V_1=-\frac{1}{\rho}(4s^2-2s+1),
\end{equation}
respectively. The components of the axial-vector part all vanish,
i.e.,
\begin{equation}
A^i=0, \quad i=0,1,2,3
\end{equation}
which is due to the diagonal metric similar to the Schwarzschild
case [32].

\section{Energy-Momentum Distribution of the Levi-Civita Solutions}

In this section, we shall use the TP version of M$\ddot{o}$ller,
Einstein, Landau-Lifshitz and Bergmann-Thomson prescriptions, given
by Eqs.( 20), (11), (12) and (13) respectively, to find the
energy-momentum distribution of the Levi-Civita solutions.

\subsection{M$\ddot{o}$ller Prescription}

When we multiply Eq.(27) by $g^{11}$ , it turns out
\begin{equation}
V^1=(4s^2-2s+1)\rho^{-8s^2+4s-1}.
\end{equation}
The non-vanishing components of the contorsion tensor, in
contravariant form, become
\begin{eqnarray}
K^{010}&=&2s\rho^{-8s^2-1}=-K^{100},\nonumber\\
K^{212}&=&\frac{1}{\alpha^2}(2s-1)\rho^{-8s^2+8s-3}=-K^{122},
\nonumber\\
K^{313}&=&2s(1-2s)\rho^{-16s^2+8s-1}=-K^{133}.
\end{eqnarray}
Clearly, the contorsion tensor is antisymmetric w.r.t. its first two
indices. By substituting Eqs.(29)-(30) in Eq.(17), the required
non-vanishing components of the supperpotential in M$\ddot{o}$ller's
tetrad theory are
\begin{equation}
U_0^{01}=\frac{-\alpha}{\kappa}(4s^2-4s+1)=-U_0^{10}.
\end{equation}
Using Eq.(31) in (20), the energy-momentum density components
vanish, i.e.,
\begin{equation}
\Xi_i^0=0, \quad i=0,1,2,3.
\end{equation}
This shows that both energy and momentum become constant in
M$\ddot{o}$ller's tetrad theory which implies that M$\ddot{o}$ller
energy-momentum distribution is independent of the coupling constant
$\lambda$.

\subsection{Einstein Prescription}

For Einstein, Landau-Lifshitz and Bergmann-Thomson prescriptions, it
is necessary to use Cartesian coordinate system to get meaningful
results. When we transform the line element (21) into Cartesian
coordinates, we obtain
\begin{equation}
ds^2=\rho^{4s}dt^2-\rho^{4s(2s-1)}\{(\frac{xdx+ydy}{\rho})^2+dz^2\}-
\alpha^2\rho^{2(1-2s)}(\frac{xdy-ydx}{\rho^2})^2.
\end{equation}
The tetrad field corresponding to this metric is
\begin{equation}
{h^a}_\mu=\left\lbrack\matrix {\rho^{2s}  &&&   0    &&&   0 &&& 0
\cr 0 &&& x \rho^{4s^2-2s-1} &&& y \rho^{4s^2-2s-1} &&&   0 \cr 0
&&& \alpha y \rho^{-2s-1} &&& -\alpha x \rho^{-2s-1}&&& 0 \cr 0
&&& 0 &&& 0 &&& \rho^{2s(2s-1)} \cr } \right\rbrack
\end{equation}
with its inverse
\begin{equation}
{h_a}^\mu=\left\lbrack\matrix {\rho^{-2s}  &&&   0    &&&   0 &&& 0
\cr 0        &&& \frac{x}{x^2-y^2}\rho^{-4s^2+2s+1}
&&&\frac{y}{y^2-x^2}\rho^{-4s^2+2s+1} &&& 0 \cr 0 &&&
\frac{y}{\alpha(y^2-x^2)}\rho^{1+2s} &&&
-\frac{x}{\alpha(x^2-y^2)}\rho^{1+2s}&&& 0 \cr 0 &&& 0 &&& 0 &&&
\rho^{2s(1-2s)} \cr } \right\rbrack.
\end{equation}
The non-vanishing components of the Weitzenb$\ddot{o}$ck connection
are
\begin{eqnarray} {\Gamma^0}_{01}&=&\frac{2sx}{\rho^2},\quad
{\Gamma^0}_{02}=\frac{2sy}{\rho^2},\nonumber\\
{\Gamma^1}_{11}&=&\frac{2sx}{\rho^4}(2sx^2-\rho^2),\quad
{\Gamma^2}_{22}=\frac{2sy}{\rho^4}(2sy^2-\rho^2),\nonumber\\
{\Gamma^1}_{12}&=&\frac{2sy}{\rho^4}(2sx^2-\rho^2),\quad
{\Gamma^2}_{21}=\frac{2sx}{\rho^4}(2sy^2-\rho^2), \nonumber\\
{\Gamma^1}_{21}&=&\frac{y}{\rho^4}(4s^2x^2-\rho^2),\quad
{\Gamma^2}_{12}=\frac{x}{\rho^4}(4s^2y^2-\rho^2), \nonumber\\
{\Gamma^1}_{22}&=&\frac{x}{\rho^4}(4s^2y^2+\rho^2),\quad
{\Gamma^2}_{11}=\frac{y}{\rho^4}(4s^2x^2+\rho^2), \nonumber\\
{\Gamma^3}_{31}&=&\frac{2sx}{\rho^2}(2s-1), \quad
{\Gamma^3}_{31}=\frac{2sy}{\rho^2}(2s-1)
\end{eqnarray}
and the components of the torsion tensor, in contravariant form, are
\begin{eqnarray}
T^{001}&=& 2sx\rho^{-8s^2-2}=-T^{010},\nonumber\\
T^{002}&=&2sy\rho^{-8s^2-2}=-T^{020}, \nonumber\\
T^{112}&=&-\alpha^{-2}y(1-2s)\rho^{-8s^2+8s-2}=-T^{121},\nonumber\\
T^{221}&=&-\alpha^{-2}x(1-2s)\rho^{-8s^2+8s-2}=-T^{212},\nonumber\\
T^{331}&=&2sx(1-2s)\rho^{-16s^2+8s-2}=-T^{313},\nonumber\\
T^{332}&=&2sy(1-2s)\rho^{-16s^2+8s-2}=-T^{323}.
\end{eqnarray}
Using Eqs.(16) and (37) in Eq.(15), the required non-vanishing
components of the tensor $S^{\mu\nu\lambda}$, in mixed form, are
\begin{eqnarray}
{S_0}^{01}=-x(4s^2-10s+1)\rho^{-8s^2+4s-2},\\
{S_0}^{02}=-y(4s^2-10s+1)\rho^{-8s^2+4s-2}.
\end{eqnarray}
When we make use of these values and $h=\alpha\rho^{8s^2-4s}$ in
Eq.(14), the non-zero components of the Freud's superpotential turn
out to be
\begin{eqnarray}
{U_0}^{01}=-\alpha x (4s^2-10s+1)\rho^{-2},\\
{U_0}^{02}=-\alpha y (4s^2-10s+1)\rho^{-2}.
\end{eqnarray}
Using Eqs.(40)-(41) in Eq.(11), the components of energy-momentum
density become
\begin{equation}
hE_\mu^0=0, \quad \mu=0,1,2,3
\end{equation}
which gives constant energy-momentum.

\subsection{Landau-Lifshitz Prescription}

When we use Eqs.(40)-(41) and the values of $h$ and $g^{\mu\nu}$ in
Eq.(12), the components of energy-momentum density in this
prescription become
\begin{eqnarray}
hL^{00}&=&-\frac{2\alpha^2s}{\pi}(4s^3-14s^2+11s-1)\rho^{8s^2-8s-2},\\
hL^{0i}&=&0, \quad i=1,2,3.
\end{eqnarray}
Here momentum also becomes constant.

\subsection{Bergmann-Thomson Prescription}

Now we replace Eqs.(40)-(41) and the values of $g^{\mu\nu}$ in
Eq.(13), so that the components of energy-momentum density in
Bergmann-Thomson prescription become
\begin{eqnarray}
hB^{00}&=&\frac{\alpha s}{\pi}(4s^2-10s+1)\rho^{-4s-2},\\
hB^{0i}&=&0, \quad i=1,2,3.
\end{eqnarray}
Here again we have constant momentum.

\section{Summary and Discussion}

The debate of the localization of energy-momentum has been an open
issue since the time of Einstein when he formulated the well-known
relation between mass and energy. Misner et al. [1] concluded that
energy can only be localized in spherical coordinates. But, soon
after, Cooperstock and Sarracino [43] demonstrated that if the
energy is localizable in spherical systems then it can be
localized in any system. Bondi [44] rejected the idea of
non-localization of energy in GR due to the reason that there
should be some form of energy which contributes to gravitation and
hence its location can, in principle, be found. Many authors
believed that a tetrad theory should describe more than a pure
gravitation field [45]. In fact, M$\ddot{o}$ller [46] considered
this possibility in his earlier attempt to modify GR.

This paper continues the investigation for the following two issues:
Firstly, we find the TP version of the Levi-Civita vacuum solutions
and evaluate the irreducible parts of the torsion tensor. Secondly,
we evaluate the energy-momentum density components of the
Levi-Civita vacuum solutions by using the TP version of
M$\ddot{o}$ller, Einstein, Landau-Lifshitz and Bergmann-Thomson
prescriptions. The axial-vector torsion vanishes because the metric
is diagonal similar to the case of Schwarzschild metric [32]. The
energy-momentum distributions for each prescription are given in the
following table:
\vspace{0.5cm}

{\bf {\small Table:}} {\small \textbf{Energy-Momentum Densities of
the Levi-Civita Metric in TP}}

\vspace{0.5cm}

\begin{center}
\begin{tabular}{|c|c|c|}
\hline{\bf Prescription}&{\bf Energy Density}&{\bf Mom. Density}\\
\hline M$\ddot{o}$ller & $ \Xi_0^0=0$ & $\Xi_i^0=0$ \\
\hline Einstein & $ hE_0^0=0$& $hE_i^0=0 $ \\ \hline
Landau-Lifshitz & $
hL^{00}=-\frac{2\alpha^2s}{\pi}(4s^3-14s^2+11s-1)\rho^{8s^2-8s-2}$
& $hL^{0i}=0
$ \\
\hline Bergmann & $ hB^{00}=\frac{\alpha
s}{\pi}(4s^2-10s+1)\rho^{-4s-2}$ &
$hB^{0i}=0$ \\
\hline
\end{tabular}
\end{center}

These results show that momentum is constant in each prescription
which coincides with the results of GR [18]. Further, energy density
becomes similar for both the theories only in the case of
M$\ddot{o}$ller prescription. However, energy density is different
in the remaining prescriptions and do not match with the
corresponding densities in GR. We also note that M$\ddot{o}$ller
energy-momentum distribution is independent of the coupling constant
$\lambda$ in TPG. It is worth mentioning here that energy-momentum
becomes constant both in GR and TPT when we choose $s=0$, as
expected for Minkowski spacetime.

\vspace{0.5cm}


{\bf Acknowledgment}

\vspace{0.5cm}

We appreciate the financial assistance of the HEC during this work.
\vspace{0.5cm}

{\bf References}

\begin{description}

\item{[1]} Misner, C.W., Thorne, K.S. and Wheeler, J.A.:
           \textit{Gravitation} (Freeman, New York, 1973).

\item{[21]} Einstein, A.: Sitzungsber. Preus. Akad. Wiss. Berlin (Math. Phys.)
            \textbf{778}(1915), Addendum ibid \textbf{779}(1915).

\item{[3]} Landau, L.D. and Lifshitz, E.M.: \textit{The Classical Theory
           of Fields} (Addison-Wesley Press, New York, 1962).

\item{[4]} Papapetrou, A.: \textit{Proc. R. Irish Acad. } \textbf{A52}(1948)11.

\item{[5]} Bergmann, P.G. and Thomson, R.: Phys. Rev.
           \textbf{89}(1958)400.

\item{[6]} Tolman, R.C.: \textit{Relativity, Thermodynamics and
           Cosmology} (Oxford University Press, Oxford, 1934).

\item{[7]} Weinberg, S.: \textit{Gravitation and Cosmology} (Wiley, New
           York, 1972).

\item{[8]} M$\ddot{o}$ller, C.: Ann. Phys. (N.Y.) \textbf{4}(1958)347.

\item{[9]} Virbhadra, K.S.: Phys. Rev. \textbf{D41}(1990)1086.

\item{[10]} Virbhadra, K.S.: Phys. Rev. \textbf{D42}(1990)2919.

\item{[11]} Rosen, N. and Virbhadra, K.S.: Gen. Rel. Grav.  \textbf{25}(1993)429.

\item{[12]} Chamorro, A. and Virbhadra, K.S.: Pramana J. Phys. \textbf{45}(1995)181.

\item{[13]} Aguirregabiria, J.M., Chamorro, A. and Virbhadra, K.S.: Gen. Rel.
Grav. \textbf{28}(1996)1393.

\item{[14]} Virbhadra, K.S.: Phys. Rev. \textbf{D60}(1999)104041.

\item{[15]} Penrose, R.: \textit{Proc. Roy. Soc., London }\textbf{A381}(1982)53.

\item{[16]} Tod, K.P.: \textit{Proc. Roy. Soc., London }\textbf{A388}(1983)457.

\item{[17]} Sharif, M.: Int. J. Mod. Phys. \textbf{A17}(2002)1175;
            \textit{ibid} \textbf{A18}(2003)4361; \textbf{A19}(2004)1495;
            \textbf{D13}(2004)1019;\\
            Sharif, M. and Fatima, T.: Nouvo Cim. \textbf{B120}(2005)533;\\
            Xulu, S.S.: Astrophys. Space Sci. \textbf{283}(2003)23.

\item{[18]} Sharif, M.: Braz. J. Phys. \textbf{37}(2007)1292.

\item{[19]} Einstein, A.: Sitzungsber. Preuss. Akad. Wiss.
            (1928)217.

\item{[20]} Hayashi, K. and Nakano, T.: Prog. Theor. Phys. {\bf 38}(1967)491.

\item{[21]} Hehl, F.W., McCrea, J.D., Mielke, E.W. and Ne'emann, Y.: Phys.
           Rep. {\bf 258}(1995)1.

\item{[22]} Hayashi, K. and Shirafuji, T.: Phys. Rev. {\bf D19}(1979)3524.

\item{[23]} Weitzenb$\ddot{o}$ck, R.: {\it Invarianten Theorie}
           (Gronningen: Noordhoft, 1923).

\item{[24]} De Andrade, V.C. and Pereira,  J.G.: Phys. Rev. {\bf
           D56}(1997)4689.

\item{[25]} Gronwald, F. and Hehl, F.W.: {\it Proceedings of the
           School of Cosmology and Gravitation on Quantum Gravity},
           eds. Bergmann, P.G., De Sabata, V. and  Tender, H.J.(World Scientific, 1995);\\
           Blagojecvic, M. {\it Gravitation and Gauge Symmetries} (IOP
           publishing, 2002);\\
           Hammond, R.T.: Rep. Prog. Phys. {\bf 65}(2002)599.

\item{[26]} Gronwald, F. and Hehl, F.W.: {\it On the Gauge Aspects of Gravity,
           Proceedings of the 14th School of Cosmology and Gravitation},
           eds. Bergmann, P.G., De Sabata, V. and  Tender, H.J.(World Scientific, 1996).

\item{[27]} Mikhail, F.I., Wanas, M.I., Hindawi, A. and Lashin, E.I.: Int. J. Theo.
            Phys. \textbf{32}(1993)1627.

\item{[28]} Hehl, F.W. and Macias, A.: Int. J. Mod. Phys. {\bf
D8}(1999)399;\\
            Obukhov, Yu N., Vlachynsky, E.J., Esser, W.,
            Tresguerres, R. and Hehl, F.W.: Phys. Lett. {\bf A220}(1996)1;\\
            Baekler, P., Gurses, M., Hehl, F.W. and McCrea, J.D.: Phys.
            Lett. {\bf A128}(1988)245;\\
            Vlachynsky, E.J. Esser, W., Tresguerres, R. and Hehl, F.W.:
            Class. Quantum Grav. {\bf 13}(1996)3253;\\
            Ho, J.K., Chern, D.C. and Nester, J.M.: Chin. J. Phys.
            {\bf 35}(1997)640;\\
            Hehl, F.W., Lord, E.A. and Smally, L.L.: Gen. Rel.
            Grav. {\bf13}(1981)1037;\\
            Kawa, T. and Toma, N.: Prog. Theor. Phys.
            {\bf 87}(1992)583;

\item{[29]} Vargas, T.: Gen. Rel. Grav. \textbf{30}(2004)1255.

\item{[30]} Rosen, N.: Gen. Rel. Grav. \textbf{26}(1994)323.

\item{[31]} Sharif, M. and Nazir, Kanwal: Braz. J. Phys. 38(2008)156;
\emph{Energy-Momentum Distribution in General Relativity and
Teleparallel Theory of Gravitation}, accepted for publication in
Canadian J. Phys;\\
Nashed, G.G.L.: Phys. Rev. \textbf{D66}(2002)060415; Gen. Rel. Grav.
\textbf{34}(2002)1074.

\item{[32]} Pereira, J.G., Vargas, T. and Zhang, C.M.: Class. Quantum Grav.
            {\bf 18}(2001)833.

\item{[33]} Sharif, M. and Amir, M. Jamil.: Gen. Rel. Grav. \textbf{38}(2006)1735.

\item{[34]} Sharif, M. and Amir, M. Jamil.: Mod. Phys. Lett. \textbf{A22}(2007)425.

\item{[35]} Sharif, M. and Amir, M. Jamil.: Gen. Rel. Grav. \textbf{39}(2007)989.

\item{[36]} Sharif, M. and Amir, M. Jamil.: \emph{Teleparallel Energy-Momentum
Distribution of Static Axially Symmetric Spacetimes}, accepted for
publication in Mod. Phys. Lett. \textbf{A}.

\item{[37]} Aldrovendi, R. and Pereira, J.G.: {\it An Introduction to
            Gravitation Theory} (preprint).

\item{[38]} Aldrovandi and Pereira, J.G.: {\it An Introduction to
            Geometrical Physics} (World Scientific, 1995).

\item{[39]} Kramer, D., Stephani, H., Hearlt, E. and MacCallum, M.A.H.:
            \textit{ Exact Solution of Einstein's Field Equations} (Cambridge
            University Press, 2003).

\item{[40]} Bicak, J., Ledivinka, T., Schmidt, B.G. and Zofka, M.:
            Class. Quantum Grav. \textbf{21}(2004)1583.

\item{[41]} da Silva, M.F.A, Herrera, L., Paiva, M.F. and Santos, N.O.: J. Math.
            Phys. {\bf 36}(1995)3625.

\item{[42]} Herrera, L., Paiva, M.F. and Santos, N.O.: J. Math.
            Phys. {\bf 40}(1999)4064.

\item{[43]} Cooperstock, F.I. and Sarracino, R.S.: J. Phys. A: Math.
            Gen. \textbf{11}(1978)877.

\item{[44]} Bondi, H.: \textit{ Proc. R. Soc. London } \textbf{A427}(1990)249.

\item{[45]} Nashed, G.G.L.: Nouvo Cim. \textbf{B117}(2002)521.

\item{[46]} M$\ddot{o}$ller, C.: Mat. Fys. Medd. Dan. Vid. Selsk. \textbf{1}(1961)10.
\end{description}
\end{document}